\documentclass[11pt]{article}
\usepackage{hyperref,xcolor}
\usepackage{cite}
\usepackage{amsmath}
\usepackage{lineno}
\usepackage{amssymb}
\usepackage{footnote}
\makesavenoteenv{minipage}
\newcommand{\nl}{\\ \nonumber  &&}
\newcommand{\mc}{\mathcal}

\renewcommand{\title}[1]{%
	\bigskip%
	\begin{center}%
		\Large\bf #1%
	\end{center}%
	\vskip .2in}

\renewcommand{\author}[1]{%
	{\begin{center}
			#1
\end{center}}}
\newcommand{\address}[1]{\vspace{-1.7em}\vspace{0pt}
	{\begin{center}
			\it #1
\end{center}}}

\begin{document}
	\begin{titlepage}
		\title{Ghosts in higher derivative Maxwell-Chern-Simon's theory and $\mc{PT}-$symmetry 
		     }
		
		\author
		{
			Biswajit Paul  $\,^{\rm a,b}$,
			Himangshu Dhar $\,^{\rm a,c}$,			
			Biswajit Saha    $\,^{\rm a, e}$
		  } 
		\address{$^{\rm a}$ National Institute of Technology Agartala \\
			Jirania, Tripura -799 055, India }\footnote{
			{$^{\rm b}$\tt biswajit.thep@gmail.com,}
			{$^{\rm c}$\tt himangshu171@gmail.com,}
			{$^{\rm e}$\tt biswajit.physics@gmail.com}}
		
		\begin{abstract}
Ghost fields  in quantum field theory have been a long standing problem. Specifically, theories with higher derivatives involve ghosts that appear in the Hamiltonian in the form of linear momenta term, which is commonly known as the Ostrogradski ghost. Higher derivative theories may involve both types of constraints i.e. first class and second class. Interestingly, these higher derivative theories may have non-Hermitian Hamiltonian respecting $\mc{PT}-$symmetries. In this paper, we have considered the $\mc{PT}-$symmetric nature of the extended Maxwell-Chern-Simon's theory and employed the second class constraints to remove the linear momenta terms causing the instabilities. We found that the removal is not complete rather conditions arise among the coefficients of the operator ${Q}$.
		\end{abstract}
	\end{titlepage} 
\section{INTRODUCTION}
We know from usual quantum mechanics  that for  real energy eigenvalues of a quantum theory, it is required that we must have a Hermitian Hamiltonian i.e. $H = H^\dagger$. Hermiticity has been the one and only condition for the reality of the quatum theory. The concept of $\mc{PT}$-symmetry, which mean reflections in space and time separately, is not new in physics.  But recently Bender et al. has pointed out that $\mc{PT}$ symmetry of the Hamiltonian i.e. $H = H^{\mc{PT}}$ can be a very important ingredient while analysing the eigen values of a theory which has a non-Hermitian Hamiltonian \cite{bender1998,bender1998gh}. According to this, for a non-Hermitian Hamiltonian, if it has unbroken $\mc{PT}-$symmetry then  the energy spectrum of the theory can be real. This means, now more systems which were earlier outright thought to have complex eigen values due to lack of Hermiticity, can  be added and analysed to have real eigen values. 
Another requirement for an acceptable quantum theory is that there should be a lower bound in the Hamiltonian otherwise it can give rise to infinite negative enegy states. These `negative norm' states are known as ghost states and removal of these ghost states is the utmost necessity to build up the corresponding quantum interpretation.  It is seen that Hamiltonian corresponding to the  higher derivative theories poses ghosts and hence these are not bounded from below. This motivated us to keep on searching for the non-Hermitian Hamiltonians (especially belonging to the higher derivative theories) and check if they really are bounded from below.  

 $\mc{PT}-$symmetries have very interesting features. Like, the parity operator $\mc{P}$ is a linear operator while $\mc{T}$ is antilinear operator, both being involutory. In this case the unitarity i.e. preserving the probability density is defined in a new way, with  an additional help from the operator, called $\mc{C}$-operator\cite{bender2004}. The $\mc{C}$ operator commutes with both the $\mc{PT} $ and the Hamiltonian operator. Since its introduction,  the use of  $\mc{PT}$ symmetric nature have been used in numerrous fields
\cite{bender1998,bender2008, bender2004sv }. As in \cite{alexandre2020}, supersymmetric non-Hermitian Hamiltonians were considered with $\mc{PT}-$symmetry and mass splitings between boson and fermion were found out which offered a novel non-Hermitian mechanism for soft supersymmetry breaking. From the experimental perspective, a class of problems dealing with loss-gain systems were analysed using the $\mc{PT}-$symmetries e.g. whispering-gallery microcavities \cite{peng2014},  magnetic metamaterial \cite{lazarides2013},  array of coupled waveguides \cite{liang2014, kalozoumis2016} etc.
	
	Higher derivative(HD) theories have been a very exciting field in theoretical physics for a  very long time with recent applications in  general relativity\cite{bertolami2007,sotiriou2010, ottinger2020, hindawi1995,capozziello2008}, cosmology \cite{calcagni2006,carloni2019, evans2007, goswami2008,clifton2007}, inflationary models\cite{berkin1990,maroto1997,maeda1987, meng2004,nojiri2006}, string theory \cite{peeters2001,becker2010,bergshoeff2012,howe2003} etc. Usually, the higher derivative terms are added to the Lagrangian as a kind of perturbation term  or quantum corrections \cite{hyakutake2006,borges2019}. Due to this additional term, the HD  theory may become renormalizable \cite{stelle1977}. The canonical formulation of the HD theory is governed by the very famous Ostrogradski method \cite{ostro}. According to the Ostrodradski method, in HD theories,  velocities of the fields are also considered as independent fileds and consequently, the Hamiltonian poses linear momenta conjugate to these fields. This is very unusual for the theories where  the equation of motion is mostly second order in time\cite{ostro}. Due to this, Hamiltonian of the HD theories poses linear momenta terms. These linear momenta terms are connected to the fields where one considers the higher time derivatives of the fields as independent fields. Therefore, when one considers the quantum version of the theory, there appears negative norms and correspondingly the energy spectrum becomes unbounded. 
	
	 The Ostrogradski theorem states that the HD theories, to be free from ghosts, must be non-degenerate \cite{woodard2007, chen2013}.  The kinetic matrix of these HD theories is, in general, non-invertible and hence  constraint structure appears.  Generally constraint theories are dealt with the Dirac's constraint analysis and consequently one need to consider the reduced phase space for quantisation. In \cite{paul2017} the author has  removed the ghost from the degenerate gravity theories, with a surface term making it HD model, by solving the second class constraints. While  in the case of the Gallilean invariant Chern-Simon's model, the constraint structure was not much helpful \cite{paul2019}. The second class constraints can be useful only if the ghost field is contained in it and is solvable.  In case of the scalar-tensor theories in \cite{langlois2016}, the authors considered an HD extension of the Horndeski theory and found the condition for degeneracy. There are a class of scalar-tensor theories proposed in \cite{gao2019} which can be made ghost free by incorporating non-dynamical degrees of freedom. In fact the cosmological perturbations  can be made free from ghosts and the equations become  second order  under some choice of external physically viable functions\cite{kase2018}.  For analytic mechanics, involving arbitrary higher derivative conditions,  the ghosts were removed by incorporating degeneracy conditions in the theory \cite{motohashi2018}. Apart from these there are many attempts to solve this ghost problem  \cite{klein2016, crisostomi2017,raidal2017,crisostomi2018}. Unfortunately, we can say that to remove the ghost fields from the degenerate HD theories, there is no sure shot prescription avaiable in the literature.

 In this paper, we consider the HD Maxwell-Chern-Simon's model, which is the field theoretic extension of the earlier model introduced by Deser \& Jackiw \cite{deser1999}. The Maxwell term is parity invariant but the Chern-Simon's term  is parity-odd which makes the whole theory become parity non-invariant.  However, proper choice of transformation of the fields can make the theory $\mc{PT}-$invariant. Apart from this, as shown in \cite{deser1999}, the model posses two ghost fields one massive and another massless. Therefore we choose this model to see if we can remove these ghost fields using the help of its $\mc{PT}-$symmetric nature. 

The plan of the paper is as follows. In Sec II we describe the HD Maxwell-Chern-Simon's model and Hamiltonian is carried out using the first order formalism. In Sec III we  show the $\mc{PT}-$ trannsformation of the fields and the canonical Hamiltonian which still have the ghost fields. In Sec IV we  calculate the ghost free Hamiltonian with suitable conditions which is our original contribution. Finally, we conclude with Sec V. 
\section{MAXWELL CHERN SIMON'S MODEL}
The MCS model model with a topological mass term $m$, in 2+1 dimensions, is given by \cite{deser1999}
\begin{equation}
S = \int d^3x \Big(-\frac{1}{4} F_{\mu \nu}F^{\mu \nu} + \frac{m}{2}\epsilon^{\alpha \beta \gamma} \partial^{\rho}\partial_{\rho} A_{\alpha} \partial_{\beta} A_{\gamma}\Big).
\label{action1}
\end{equation}

The action (\ref{action1}) is metric independent and $A_\mu$ are the vector fields describing the Maxwell equations. This model originates by extending the D'Alembertian operator to the  Chern-Simon's term  and added to the Maxwell's action  which is the minimally possible combination maintaining other properties of the Chern-Simon's theory in $D=3$. $m$ has the dimension of $L^{-1}$. From action (\ref{action1}) we can deduce the Euler-Lagrange equation of motion, which is
\begin{eqnarray}
\partial_{\mu}F^{\mu\nu} + m\epsilon^{\nu\mu\rho}\partial^{\alpha}\partial_{\alpha}F_{\mu\rho}=0.
\end{eqnarray}
The above equation contains third order time derivative of the fields and hence equation of motions are higher derivative in nature. These class of theories need to be addressed differently as laid down by the Ostrogradski method \cite{ostro}. For this purpose we rename these variables  as 
\begin{eqnarray}
A_{\mu} = \xi_{1\mu}, \\
\dot{A}_{\mu} = \xi_{2\mu}.
\end{eqnarray}
Where a 'dot' indicates derivative with respect to time. As this is a 2+1 D model, greek indices run from 0 to 2 and latin indices run from 1 to 2.
In terms of the new variable $\xi$, the MCS Lagrangian from the (\ref{action1}) can be written as
\begin{eqnarray}
\nonumber
\mathcal{L} &=& \frac{1}{2}\Big( \xi_{2i}\xi_{2i} + \partial_{i}\xi_{10}\partial_{i}\xi_{10} - \partial_{i}\xi_{1j}\partial_{i}\xi_{1j} +\partial_{i}\xi_{1j}\partial_{j}\xi_{1i} - 2\xi_{2i} \partial_{i}\xi_{10} \Big) + \frac{m}{2}\epsilon_{ij}(-\dot{\xi}_{20}+ \nabla^2 \xi_{10})\partial_{i}\xi_{1j} \\  && 
- \frac{m}{2}\epsilon_{ij}(-\dot{\xi}_{2i} + \nabla^2\xi_{1i})\xi_{2j} + \frac{m}{2}\epsilon_{ij}(- \dot{\xi}_{2i} + \nabla^2 \xi_{1i})\partial_{j}\xi_{10} + \xi_{0\mu}(\xi_{2\mu} - \dot{\xi}_{1\mu}).
\label{lag1}
\end{eqnarray}
Here these newly introduced $\xi_{0\mu}$ are the Lagrange multipliers incorporated owing to the constraint relationship
\begin{equation}
\xi_{2\mu} = \dot{\xi}_{1\mu},
\end{equation}
which was reqired to convert this higher derivative system into a first order system. These Lagrange multipliers do not effect  the dynamics of the system and hence, can be removed at the later stage based on the constraint structure. The momenta of the lagrangian (\ref{lag1}) are
\begin{eqnarray}
\Pi_{\alpha\mu} = \frac{\partial \mathcal{L}}{\partial \dot{\xi}_{\alpha \mu}}.
\end{eqnarray}
We immediately get the following primary constraints 
\begin{eqnarray}
\Phi_{0\mu}&=&\Pi_{0\mu} \approx 0, \ \ \ \  \Phi_{1\mu} = \Pi_{1\mu}+\xi_{0\mu} \approx 0, \\ \nonumber 
\Phi_{20} &=& \Pi_{20} + \frac{m}{2}\epsilon_{ij}\partial_{i}\xi_{1j} \approx 0, \\ \nonumber 
\Phi_{2i} &=& \Pi_{2i} - \frac{m}{2} \epsilon_{ij}\xi_{2j} + \frac{m}{2}\epsilon_{ij}\partial_{j}\xi_{10} \approx 0. \label{p_cons}
\end{eqnarray}
All of these  are primary constraints, however, secondary contraints will emerge out once we consider their time evolution using the  Hamiltonian formulation. The canonical Hamiltonian density is defined in the usual way as
\begin{eqnarray}
\nonumber
\mathcal{H}_{can} &=&  -\frac{1}{2}\Big( \xi_{2i}\xi_{2i} + \partial_{i}\xi_{10}\partial_{i}\xi_{10} - \partial_{i}\xi_{1j}\partial_{i}\xi_{1j} +\partial_{i}\xi_{1j}\partial_{j}\xi_{1i} - 2\xi_{2i} \partial_{i}\xi_{10} \Big) - \frac{m}{2}\epsilon_{ij}\nabla^2 \xi_{10}\partial_{i}\xi_{1j} \\  && 
+ \frac{m}{2}\epsilon_{ij}\nabla^2\xi_{1i}\xi_{2j} - \frac{m}{2}\epsilon_{ij} \nabla^2\xi_{1i}\partial_{j}\xi_{10} - \xi_{0\mu}\xi_{2\mu}.
\label{canham1} 
\end{eqnarray}
The total Hamiltonian is defined as the linear combination of all the primary constraints to the canonical Hamiltonian which is given by
\begin{eqnarray}
\mathcal{H}_{T} = \int d^2x (\mathcal{H}_{can}+ \Lambda_{0\mu}\Phi_{0\mu}+\Lambda_{1\mu}\Phi_{1\mu} + \Lambda_{2\mu}\Phi_{2\mu}).
\end{eqnarray}
Here $\Lambda$'s are the Lagrange multipliers which can be determined once we consider the time evolution of the primary constraints. Please refer to \cite{mukherjee2012} for a detailed analysis of the constraint structure of this MCS model and \cite{dirac} to explore more on constraint analysis. The Poission bracket of the primary constraints with the Hamiltonian gives us the time evolution and equating them to be zero we get the following secondary constraints
\begin{eqnarray}
\Psi_{1} &=& \xi_{00} + \frac{m}{2}\epsilon_{ij} \partial_{i}\xi_{2j} \approx 0, \\
\Psi_{2} &=& \partial_{i}\xi_{0i} - \frac{m}{2}\epsilon_{ij}\nabla^2\partial_{i}\xi_{1j} \approx 0. \label{s_cons} 
\end{eqnarray}
No more constraints emerge out once we consider the time preservation of the secondary constraints. Rather, they become identically zero. These constraints are further classified as first class and second class constraints based on the poission brackets among themselves. If all the Poission brackets of a constraint are zero including itself then it is called a first class constraint else second class constraint. On this basis  $\{ \Phi_{20}, \Psi_1, \Psi_2 \}$ are first class while $\Phi_{2i}$ are second class constraints. Evidently,  there is only one primary first class constraints $\Phi_{20}$ which indicates existence of one gauge symmetry.  This higher derivative MCS  Lagrangian is invariant under the transformations
$\xi_{1\mu}' \rightarrow \xi_{1\mu} + \partial_{\mu}\lambda$, where $\lambda$ is arbitrary parameter indicating that the theory has a underlying symmetry of $U(1)$ group.
\subsection*{The reduced phase space:}
To quantise, the first class constraints will be treated as auxilliary conditions which annihilate the states. On the other hand the second class constraints are treated as operator identities among the phase space variables which helped us to reduce the phase space by  replacing the phase space variables algebraically. Effectively we have to concentrate on the canonical Hamiltonian, which after replacement of the auxilliary variable $\xi_{0\mu}$ from the constraint $\Phi_{1\mu}$ become
 
\begin{eqnarray}
\nonumber
\mathcal{H}_{can} &=&  -\frac{1}{2}\Big( \xi_{2i}\xi_{2i} + \partial_{i}\xi_{10}\partial_{i}\xi_{10} - \partial_{i}\xi_{1j}\partial_{i}\xi_{1j} +\partial_{i}\xi_{1j}\partial_{j}\xi_{1i} - 2\xi_{2i} \partial_{i}\xi_{10} \Big) - \frac{m}{2}\epsilon_{ij}\nabla^2 \xi_{10}\partial_{i}\xi_{1j} \\  && 
+ \frac{m}{2}\epsilon_{ij}\nabla^2\xi_{1i}\xi_{2j} - \frac{m}{2}\epsilon_{ij} \nabla^2\xi_{1i}\partial_{j}\xi_{10} + \Pi_{1\mu}\xi_{2\mu}.
\label{canham2} 
\end{eqnarray}
Interestingly, the canonical Hamiltonian  involves the linear momenta $\Pi_{1\mu}$ which indicates that the states defined in the Hilbert space will give  negative norms once one consider the quantum picture. In the following sections we shall try to remove this ghost field adopting the $\mc{PT}$ nature of the system.
 \section{ $\mc{PT}$ TRANSFORMATION OF FIELD VARIABLES}
 In this section we shall consider the aspect of $\mathcal{PT}$-symmetric theories and their effect when we consider the MCS action. 

 Being a Chern-Simon class of theory, under the partiy, the fields transform as 
 \begin{eqnarray}
 \mc{P} \xi_{10}(t,r) \mc{P}^{-1} &=& \xi_{10}(t,r'),\\
 \mc{P}\xi_{11}(t,r)\mc{P}^{-1} &=& -\xi_{11}(t,r'),\\
 \mc{P}\xi_{12}(t,r)\mc{P}^{-1} &=& \xi_{12}(t,r'),\\
 \mc{P} \xi_{20}(t,r) \mc{P}^{-1} &=& \xi_{20}(t,r'),\\
 \mc{P}\xi_{21}(t,r)\mc{P}^{-1} &=& -\xi_{21}(t,r'),\\
 \mc{P}\xi_{22}(t,r)\mc{P}^{-1} &=& \xi_{22}(t,r'),\\
 \mc{P} i \mc{P}^{-1} &=& i.
 \end{eqnarray}

 Whereas, under the time reversal operator, the transformations of the variables  are\\
 \begin{eqnarray}
 \mc{T} \xi_{10}(t,r) \mc{T}^{-1} &=& \xi_{10}(t,r'),\\
 \mc{T}\xi_{11}(t,r)\mc{T}^{-1} &=& -\xi_{11}(t,r'),\\
 \mc{T}\xi_{12}(t,r)\mc{T}^{-1} &=& -\xi_{12}(t,r'),\\
 \mc{T} \xi_{20}(t,r) \mc{T}^{-1} &=& -\xi_{20}(t,r'),\\
 \mc{T}\xi_{21}(t,r)\mc{T}^{-1} &=& \xi_{21}(t,r'),\\
 \mc{T}\xi_{22}(t,r)\mc{T}^{-1} &=& \xi_{22}(t,r'),\\
\mc{T} i \mc{T}^{-1} &=& -i.
 \end{eqnarray}
 
Hence effect of the combined $\mc{PT}$-transformation on the variables are
\begin{eqnarray}
\mc{PT} \xi_{10}(t,r) (\mc{PT})^{-1} &=& \xi_{10}(t,r'),\\
\mc{PT}\xi_{11}(t,r)(\mc{PT})^{-1} &=& \xi_{11}(t,r'),\\
\mc{PT}\xi_{12}(t,r)(\mc{PT})^{-1} &=& -\xi_{12}(t,r'),\\
\mc{PT} \xi_{20}(t,r) (\mc{PT})^{-1} &=& -\xi_{20}(t,r'),\\
\mc{PT}\xi_{21}(t,r)(\mc{PT})^{-1} &=& -\xi_{21}(t,r'),\\
\mc{PT}\xi_{22}(t,r)(\mc{PT})^{-1} &=& \xi_{22}(t,r'),\\
\mc{PT} i (\mc{PT})^{-1} &=& -i.
\end{eqnarray}
Wih these basic definitions of transformation of the fields, it is imperative to see how the Hamiltonian behave under the $\mc{PT}$-symmetry. It is seen that the canonical  Hamiltonian {remains invariant} under the $\mc{PT}$ symmetry as
\begin{equation}
\mc{PT}\mc{H}_{can}(\mc{PT})^{-1} = \mc{H}_{can}.
\end{equation} 

\subsection*{{Reason for considering imaginary sector:}}
In this theory there are two distinct fields $\xi_{1\mu}$ and $\xi_{2\mu}$. So in the corresponding quantum picture, the creation and annihilation operators will be the function of these phase-space variables. It was shown in \cite{bender2008}, with proper argument,  that out of these fields, $\xi_{1\mu}$ possesses an unbounded sector as it has a negative Dirac norm. This issue is characteristic of the HD theories. It was shown for the Pias Ulhenbeck model that the ghost sector lies along the imaginary axis and due to this, along the real axis, the integrations blow up. So, to avoid integrations along the imaginary axis we consider a transformation keeping in view that the commutator algebra should be respected. We take a similarity transformation of the fields involving ghosts
\begin{eqnarray}
\xi_{1\mu} = i\tilde{\xi}_{1\mu}, \\ 
\Pi_{1\mu} = -i \tilde{\Pi}_{1\mu}.
\label{isospectral}
\end{eqnarray}
When we apply the transformation (\ref{isospectral}) to the Hamiltonian 
(\ref{canham2}) we get 
\begin{eqnarray}
\nonumber
\tilde{\mathcal{H}}_{can} &=&  -\frac{1}{2}\Big( \xi_{2i}\xi_{2i} -\partial_{i}\tilde{\xi}_{10}\partial_{i}\tilde{\xi}_{10} + \partial_{i}\tilde{\xi}_{1j}\partial_{i}\tilde{\xi}_{1j} -\partial_{i}\tilde{\xi}_{1j}\partial_{j}\tilde{\xi}_{1i} - i 2\tilde{\xi}_{2i} \partial_{i}\tilde{\xi}_{10} \Big) + \frac{m}{2}\epsilon_{ij}\nabla^2 \tilde{\xi}_{10}\partial_{i}\tilde{\xi}_{1j} \\  && 
+ i \frac{m}{2}\epsilon_{ij}\nabla^2 \tilde{\xi}_{1i}\xi_{2j} + \frac{m}{2}\epsilon_{ij} \nabla^2\tilde{\xi}_{1i}\partial_{j}\tilde{\xi}_{10} - i\tilde{\Pi}_{1\mu}\xi_{2\mu}.
\label{canham3} 
\end{eqnarray}
In this Hamiltonian there are two distinct part one we call $H_0$ which is the real part and the other $H_1$ which consist of the imaginary part so that we can write 
\begin{equation}
\tilde{\mc{H}} = H_0 + H_1.
\end{equation} This Hamiltonian is non-Hermitian as $\tilde{\mathcal{H}}_{can} \ne \tilde{\mathcal{H}}^\dagger_{can}$ and hence a sense may arise that the eigen values of this Hamiltonian are not real. But this is not the case here as this Hamiltonian is $\mc{PT}$-symmetric 
\begin{equation}
\mc{PT}\tilde{\mathcal{H}}_{can}(\mc{PT})^{-1} = \tilde{\mathcal{H}}_{can}. 
\end{equation}
The $\mc{PT}-$symmetric nature of this theory confirms us that it should have real energy spectrum. 
\section{SIMILARITY TRANSFORMATIONS OF THE CANONICAL HAMILTONIAN OWING TO $\mc{PT}$}
According to the $\mc{PT}-$symmetric theories the norm of a state $|n \rangle$ is defined as \cite{bender1998, bender1998gh}
\begin{equation}
\langle n| n\rangle^{\mc{PT}} = \langle n| e^{-Q}|n\rangle.
\end{equation}
It is worth noting  the new definition of the ket states. $Q$ is connected to the symmetry oprator as $\mc{C} = e^{Q}\mc{P}$ and under the redefined norm, the ghost fields can behave as usual fields. The $\mc{C}$ operator obeys the following relations
\begin{eqnarray}
\mc{C}^2 &=& 1, \\
\left[ \mc{C}, \mc{PT} \right] &=&0, \\
\left[\mc{C}, \tilde{H}_{can} \right] &=& 0.
\label{c_brackets}
\end{eqnarray}
The last relation in (\ref{c_brackets}) gives us an additional and very important relation 
\begin{equation}
e^Q \tilde{H}_{can} e^{-Q} = H_0 - H_1.
\label{c_brackets2}
\end{equation}
For the present model let us define 
$Q$ as \cite{bender1998}
\begin{equation}
Q = \alpha \tilde{\xi}_{1\mu}\xi_{1\mu} + \beta \tilde{\Pi}_{1\mu}\Pi_{2\mu}.
\end{equation}
Here the coefficients $\alpha$ and $\beta$ are  not function of the spacetime variables. However, we can put condition on their values so as to get the desired outcome.  The basic fields transform as 
\begin{eqnarray}
e^{Q}\tilde{\xi}_{1\mu}e^{-Q} &=& c \tilde{\xi}_{1\mu} + i ds \Pi_{2\mu}, \\ 
e^{Q}\xi_{2\mu}e^{-Q} &=& c \xi_{2\mu} + i ds \tilde{\Pi}_{1\mu}, \\
e^{Q}\tilde{\Pi}_{1\mu}e^{-Q} &=& c \tilde{\Pi}_{1\mu} - \frac{is}{d} \xi_{2\mu}, \\
e^{Q}\Pi_{2\mu}e^{-Q} &=& c \Pi_{2\mu} - \frac{is}{d} \tilde{\xi}_{1\mu}. \label{field_similarity}
\end{eqnarray}
Where we have taken $c = cosh\sqrt{\alpha\beta}, s=sinh\sqrt{\alpha\beta}$ and $d = \sqrt{\frac{\beta}{\alpha}}$. Employing the similarity transforms of the basic fields  in (\ref{canham3}) we get
\begin{eqnarray}
\nonumber 
e^{Q}\tilde{\mc{H}}_{can}e^{-Q} &=&  -i(c^2+s^2) \tilde{\Pi}_{10}\xi_{20} + cds \tilde{\Pi}_{10}^2 -\frac{cs}{d} \xi_{20}^2 + (cds + \frac{d^2s^2}{2}) \tilde{\Pi}_{1i}^2 -i(c^2 + s^2 +cds)\Pi_{1i}\xi_{2i} \nl+ k_3 \tilde{\Pi}_{1i}^2 -\frac{c^2d + cs}{2d} \xi_{2i}^2 -cds \tilde{\Pi}_{1i}\partial_{i}\tilde{\xi}_{10} +ic^2 \xi_{2i}\partial_{i}\tilde{\xi}_{10} 
+\frac{c^2}{2} \partial_{i} \tilde{\xi}_{10}\partial_{i} \tilde{\xi}_{10} + icds \partial_{i}\Pi_{20} \partial_{i}\tilde{\xi}_{10} \nl
-id^2 s^2 \tilde{\Pi}_{1i}\partial_{i}\Pi_{20} -cds \xi_{2i} \partial_{i}\Pi_{20} -\frac{d^2 s^2}{2} \partial_{i}\Pi_{20}\partial_{i}\Pi_{20}   +\frac{c^2}{2} \partial_{i}\tilde{\xi}_{1j}\partial_{j}\tilde{\xi}_{1i} -icds\partial_{i}\Pi_{2j}\partial_{j} \tilde{\xi}_{1i} \nl -\frac{c^2}{2} \partial_{i}\tilde{\xi}_{1j} \partial_{i}\tilde{\xi}_{1j}   -icds \partial_{i}\Pi_{2j}\partial_{i}\tilde{\xi}_{1j} - \frac{d^2 s^2}{2} \partial_{i}\Pi_{2j} \partial_j\Pi_{2i} +  \frac{d^2 s^2}{2} \partial_{i}\Pi_{2j} \partial_i\Pi_{2j} \nl + \frac{cdms}{2} \epsilon_{ij} \tilde{\Pi}_{1i}\nabla^2 \tilde{\xi}_{1j} - \frac{ic^2 m}{2} \epsilon_{ij} \xi_{2i} \nabla^2 \tilde{\xi}_{1j} - \frac{m c^2 }{2} \epsilon_{ij} \partial_{i} \tilde{\xi}_{10} \nabla^2 \tilde{\xi}_{1j} - \frac{icdms}{2} \epsilon_{ij} \partial_{i}\Pi_{20}\nabla^{2} \tilde{\xi}_{1j} \nl + \frac{imd^2 s^2}{2} \epsilon_{ij}\tilde{\Pi}_{1i}\nabla^2 \Pi_{2j} + \frac{mcds}{2} \epsilon_{ij}\xi_{2i}\nabla^2 \Pi_{2j} -\frac{i mcds}{2}\epsilon_{ij} \partial_{i} \tilde{\xi}_{10} \nabla^2 \Pi_{2j} 
+ \frac{md^2s^2}{2} \epsilon_{ij}\partial_{i}\Pi_{20} \nabla^2 \Pi_{2j}  \nl  - \frac{mc^2}{2} \epsilon_{ij}\partial_{j}\tilde{\xi}_{1i}\nabla^2 \tilde{\xi}_{10} - \frac{imcds}{2} \epsilon_{ij} \partial_{j}\tilde{\xi}_{1i}\nabla^2\Pi_{20} - \frac{imcds}{2} \epsilon_{ij}\partial_{j}\Pi_{2i}\nabla^2 \tilde{\xi}_{10} +  \frac{md^2 s^2}{2} \epsilon_{ij} \partial_{j}\Pi_{2i} \nabla^2 \Pi_{20}.
 \label{similarity1}
\end{eqnarray}

We can see that mere similarity transformation of the canonical Hamiltonian do not remove the ghost fields. 
Now we apply the second condition (\ref{c_brackets2}) and replace $\Pi_{1i}\xi_{2i}$ in (\ref{similarity1}) which,  after some algabraic simplification, takes the form 
\begin{eqnarray}
\nonumber 
e^{Q}\tilde{\mc{H}}_{can}e^{-Q} 
&=& -ik k_1 \tilde{\Pi}_{10}\xi_{20} -kds k_2 \tilde{\Pi}_{10}^2 - \frac{k s k_2}{k} \xi_{20}^2 - \frac{kds k_3}{4c} \tilde{\Pi}_{1i}^2  + \frac{k k_4}{2cd} \xi_{2i}^2 - dsk k_2 \tilde{\Pi}_{1i}\partial_{i}\tilde{\xi}_{10}   + \frac{ikk_5}{4c} \xi_{2i}\partial_{i}\tilde{\xi}_{10} \nl
+ \frac{k k_5}{8c} \partial_{i} \tilde{\xi}_{10}\partial_{i} \tilde{\xi}_{10} + ikdsk_2 \partial_{i}\Pi_{20} \partial_{i}\tilde{\xi}_{10}
+ \frac{ikd^2 s^2 k_2}{c} \tilde{\Pi}_{1i}\partial_{i}\Pi_{20} -dsk k_2\xi_{2i} \partial_{i}\Pi_{20}  - \frac{kd^2 s^2 k_2}{2c} \partial_{i}\Pi_{20}\partial_{i}\Pi_{20} \nl - \frac{kk_{5}}{8c} \partial_{i}\tilde{\xi}_{1j}\partial_{j}\tilde{\xi}_{1i} -ik d s k_2 \partial_{i}\Pi_{2j}\partial_{j} \tilde{\xi}_{1i} + \frac{kk_{5}}{8c} \partial_{i}\tilde{\xi}_{1j} \partial_{i}\tilde{\xi}_{1j}   -ikdsk_2 \partial_{i}\Pi_{2j}\partial_{i}\tilde{\xi}_{1j}  - \frac{kd^2 s^2 k_2}{2c} \partial_{i}\Pi_{2j} \partial_j\Pi_{2i} \nl +  \frac{k d^2 s^2 k_2}{2c} \partial_{i}\Pi_{2j} \partial_i\Pi_{2j} - \frac{kdmk_2}{2} \epsilon_{ij} \tilde{\Pi}_{1i}\nabla^2 \tilde{\xi}_{1j} + \frac{imk(k_5+ 4(k_2 - 1))}{8c} \epsilon_{ij} \xi_{2i} \nabla^2 \tilde{\xi}_{1j} \nl - \frac{km k_5}{8c} \epsilon_{ij} \partial_{i} \tilde{\xi}_{10} \nabla^2 \tilde{\xi}_{1j} - \frac{ikdmsk_2}{2} \partial_{i}\Pi_{20}\nabla^{2} \tilde{\xi}_{1j}  + i \frac{i k d^2 s^2 m k_2}{2c} \epsilon_{ij}\tilde{\Pi}_{1i}\nabla^2 \Pi_{2j} + \frac{kdms k_2}{2} \epsilon_{ij}\xi_{2i}\nabla^2 \Pi_{2j} \nl -\frac{ikdms k_2}{2}  \epsilon_{ij} \partial_{i} \tilde{\xi}_{10} \nabla^2 \Pi_{2j} 
+ \frac{km d^2 s^2 k_2}{2c} \epsilon_{ij}\partial_{i}\Pi_{20} \nabla^2 \Pi_{2j}    - \frac{kms k_5}{8c} \epsilon_{ij}\partial_{j}\tilde{\xi}_{1i}\nabla^2 \tilde{\xi}_{10} \\ && - \frac{ikdmsk_2}{2c} \epsilon_{ij} \partial_{j}\tilde{\xi}_{1i}\nabla^2\Pi_{20} - \frac{ikdms k_2}{2} \epsilon_{ij}\partial_{j}\Pi_{2i}\nabla^2 \tilde{\xi}_{10} +  \frac{k m d^2 s^2 k_2}{2c} \epsilon_{ij} \partial_{j}\Pi_{2i} \nabla^2 \Pi_{20}.
\label{similarity2}
\end{eqnarray} 
while the coeficients are given by 
\begin{eqnarray}
\nonumber 
&& k = \frac{1}{(2c + ds)} \nl 
k_{1} =  2 cosh\theta + 2 cosh3\theta + d sinh3\theta \nl 
k_2  = 1 + 2cosh2\theta + dsinh2\theta \nl 
k_3 = (8-d^2) cosh\theta + (4+d^2) cosh3\theta + 2d(sinh\theta + 2sinh3\theta) \nl 
k_4 = \frac{d}{2}+\frac{d}{2}cosh2\theta + dcosh4\theta + sinh2\theta + sinh4\theta + \frac{d^2}{4} sinh4\theta ) \nl 
k_5 = 4+2cosh2\theta + 2cosh4\theta + d sinh4\theta \nl 
\end{eqnarray}
\subsection*{Removing the Ghost fields}
In order to find the ghost free version of the above similarity transformed Hamiltonian (\ref{similarity2}),  we should remove the last linear momenta term $\epsilon_{ij}\tilde{\Pi}_{1i}\nabla^2\tilde{\xi}_{1j}$. This particular  momenta term can not be replaced from any of the constraints and hence a true ghost field.  However, we can find some condition to remove this term by setting its coefficient equal to zero i.e. $k_{2}=0$. This, in turn, gives us a region in the complex plane having the solution as 
\begin{eqnarray}
\theta = \frac{1}{2}\Big( 2i\pi C + log\Big[\frac{-1 \pm \sqrt{d^2 - 3}}{d+2}\Big]\Big).
\label{theta}
\end{eqnarray}
We have two solutions $\theta_{+}$ and $\theta_{-}$, for the positive and negative choices respectively. This puts a condition on our choice of d which always should be obeyed and i.e.  $d \ge \sqrt{3}$. Consequently $d+2$ is maintained to be grater than zero. In addition to this  for $\theta_{+}$  to make the numerator in (\ref{theta}) positive we have $d > 2$. So all this conditions are automatically satisfied if we consider, for $\theta_{+}$ that $d$ is greater than 2. While using the same argument it is evident that for $\theta_{-}$ we must choose $\sqrt{3} < d <2$.

Hence using the condition  $k_2 = 0$ and solving the constraint $\Psi_1$ from (\ref{s_cons}) the Hamiltonian (\ref{similarity2}) can be simplified to 
\begin{eqnarray}
\nonumber 
\mc{H}_{gf} &=& -ik k_1 \frac{m}{2}\epsilon_{ij}\partial_{i}\xi_{2j} \xi_{20} - \frac{kds k_3}{4c} \tilde{\Pi}_{1i}^2  + \frac{k k_4}{2cd} \xi_{2i}^2    + \frac{ikk_5}{4c} \xi_{2i}\partial_{i}\tilde{\xi}_{10}
+ \frac{k k_5}{8c} \partial_{i} \tilde{\xi}_{10}\partial_{i} \tilde{\xi}_{10} 
  - \frac{kk_{5}}{8c} \partial_{i}\tilde{\xi}_{1j}\partial_{j}\tilde{\xi}_{1i} \nl + \frac{kk_{5}}{8c} \partial_{i}\tilde{\xi}_{1j} \partial_{i}\tilde{\xi}_{1j}    + \frac{imk(k_5  - 4 )}{8c} \epsilon_{ij} \xi_{2i} \nabla^2 \tilde{\xi}_{1j} - \frac{km k_5}{8c} \epsilon_{ij} \partial_{i} \tilde{\xi}_{10} \nabla^2 \tilde{\xi}_{1j}   - \frac{kms k_5}{8c} \epsilon_{ij}\partial_{j}\tilde{\xi}_{1i}\nabla^2 \tilde{\xi}_{10}. 
\label{H_gf}
\end{eqnarray}
 
 It is clear from the above expression of the Hamiltonian (\ref{H_gf}) that all the terms are square powers of the respective fields and there is no term involving any linear momenta. Hence it is a bounded from below Hamiltonian and the norm will be positive and thus we have obtained the  ghost free version of the higher derivative MCS Hamiltonian under the preferred condition. 
 
\section{CONCLUSION}
Higher derivative extensions are very useful due to their own right. The MCS model discussed here is a classic example of HD theory possesing ghosts. The problem of ghosts starts back at the beginning of the quantum field theories where unwanted degrees of freedom contribute divergences and leads to instability in the theory. There have been many attempts to remove these ghost fields with different approaches while a common way is still a topic of research \cite{bender2008,klein2016,paul2019}. Keeping this in mind, we have considered the use of $\mc{PT}-$symmetries as in \cite{bender2008} where ghosts were successfully removed from the Pias-Ulhenback oscillator. However, this method is  not effective and need modifications  when the theories possess constraints. As in \cite{paul2019} the case of gallilean invariant Chern-Simon's model was considered and using the $\mc{PT}$ symmetries and the constraint structure the ghost field was removed.

But, in this paper, the constraint structure of the higher derivatinve MCS model is different as it contains both the first class and the second class contraints. We have solved the second class constraints  to remove some of the momenta. Consequently, in the further calculations, equivalent Dirac bracket should come into play instead of the Poission brackets  in the reduced phase-space. Till, there were the linear momenta terms and therefore we have applied the $\mc{PT}-$symmetric nature where the $\mc{C}$ operator was helpful. Being an additional identity, we have exploited the porperty of the  $\mc{C}-$ operator as it commutes  with both the Hamiltonian and the $\mc{PT}-$operator. This gave us the condition to replace the linear momenta term in the Hamiltonian. Another momenta, $\tilde{\Pi}_{10}$, which  was still in the Hamiltonian was removed by considering it's coefficient to be zero and the corresponding condition we have found out. The condition on the coefficiets $\alpha$ and $\beta$ shows that the ghosts were not removed for all sector. This paper thus serves us a way to exorcise the Ostrogradski ghosts from the higher derivative MCS model by incorporating both the constraint structure as well the $\mc{PT}-$symmetric nature alongwith the additional condition on the coefficients. 

\end{document}